\documentclass[twocolumn]{el-author}
\usepackage{tabularx}
\usepackage{multicol}
\usepackage{booktabs}
\usepackage{threeparttable}
\usepackage{amsmath}
\usepackage{dcolumn}
\usepackage{multirow}
\usepackage{amsfonts}
\usepackage{array}
\newcommand{\PreserveBackslash}[1]{\let\temp=\\#1\let\\=\temp}
\newcolumntype{C}[1]{>{\PreserveBackslash\centering}p{#1}}
\newcolumntype{R}[1]{>{\PreserveBackslash\raggedleft}p{#1}}
\newcolumntype{L}[1]{>{\PreserveBackslash\raggedright}p{#1}}

\newcommand{\ua}{\uparrow}
\newcommand{\nc}{\newcommand}
\nc{\da}{\downarrow} \nc{\hc}{\hat{c}} \nc{\hS}{\hat{S}}
\nc{\bra}{\langle} \nc{\ket}{\rangle} \nc{\eq}{equation (\ref}
\nc{\h}{\hat} \nc{\hT}{\h{T}}\nc{\be}{\begin{eqnarray}}
\nc{\ee}{\end{eqnarray}}\nc{\rd}{\textrm{d}}\nc{\e}{eqnarray}\nc{\hR}{\hat{R}}\nc{\Tr}{\mathrm{Tr}}
\nc{\tS}{\tilde{S}}\nc{\tr}{\mathrm{tr}}\nc{\8}{\infty}\nc{\lgs}{\bra\ua,\phi|}\nc{\rgs}{|\ua,\phi\ket}
\nc{\hU}{\hat{U}}\nc{\lfs}{\bra\phi|}\nc{\rfs}{|\phi\ket}\nc{\hZ}{\hat{Z}}\nc{\hd}{\hat{d}}\nc{\mD}{\mathcal{D}}
\nc{\bd}{\bar{d}}\nc{\bc}{\bar{c}}\nc{\mc}{\mathcal}\nc{\ea}{eqnarray}\nc{\mG}{\mathcal{G}}\nc{\bce}{\begin{center}}
\nc{\ece}{\end{center}}
\date{24th February 2014}

\begin{document}

\title{Low-complexity near-optimal signal detection for uplink large-scale MIMO systems}

\author{Xinyu Gao, Linglong Dai, Yongkui Ma, \\and Zhaocheng Wang, \emph{IET Fellow}}

\abstract{Minimum mean square error (MMSE) signal detection
algorithm is near-optimal for uplink multi-user large-scale multiple
input multiple output (MIMO) systems, but involves matrix inversion
with high complexity. In this letter, we firstly prove that the MMSE
filtering matrix for large-scale MIMO is symmetric positive
definite, based on which we propose a low-complexity near-optimal signal
detection algorithm by exploiting the Richardson method to avoid the
matrix inversion. The complexity can be reduced from ${{\cal
O}({K^3})}$ to ${{\cal O}({K^2})}$, where $K$ is the number of
users. We also provide the convergence proof of the proposed
algorithm. Simulation results show that the proposed signal
detection algorithm converges fast, and achieves the near-optimal
performance of the classical MMSE algorithm.}

\maketitle

\section{Introduction}
Large-scale multiple input multiple output (MIMO)  employing
hundreds of antennas at the base station (BS) to simultaneously
serve multiple users is a promising key technology for 5G wireless
communications \cite{1}. It can achieve orders of magnitude increase
in spectrum and energy efficiency, and one challenging issue to
realize such goal is the low-complexity signal detection algorithm
in the uplink, due to the increased dimension of large-scale MIMO
systems \cite{2}. The optimal  signal detection algorithm is the
maximum likelihood (ML) algorithm, but its complexity increases
exponentially with the number of transmit antennas, making it
impossible for large-scale MIMO. The fixed-complexity sphere
decoding (FSD) \cite{3} and tabu search (TS) \cite{4} algorithms
have been proposed with reduced complexity, but their complexity is
unfordable when the dimension of the large-scale MIMO system is
large or the modulation order is high \cite{5}. Low-complexity linear
detection algorithms such as zero-forcing (ZF) and minimum mean
square error (MMSE) with near-optimal performance have been
investigated \cite{2}, but these algorithms have to use unfavorable
matrix inversion, whose high complexity is still not acceptable for
large-scale MIMO systems. Very recently, the Neumann series
approximation algorithm has been proposed to approximate the matrix
inversion \cite{6}, which converts the matrix inversion into a
series of matrix-vector multiplications. However, only marginal
complexity reduction can be achieved.

In this letter, we propose a low-complexity near-optimal signal detection
algorithm by exploiting the Richardson method \cite{7} to avoid the
complicated matrix inversion. We firstly prove a special property of
large-scale MIMO systems that the MMSE filtering matrix is symmetric
positive definite, based on which we propose to exploit the
Richardson method to avoid the complicated matrix inversion. Then we
prove the convergence of the proposed algorithm for any initial
solution when the relaxation parameter is appropriate. Finally, we
verify through simulations that the proposed signal detection
algorithm can efficiently solve the matrix inversion problem in an
iterative way until the desired accuracy is attained, and achieve
the near-optimal performance of the MMSE algorithm with exact matrix
inversion.

\section{Large-Scale MIMO System Model}
We consider a uplink multi-user large-scale MIMO system which
employs ${N}$ antennas at the BS to simultaneously serve ${K}$
single-antenna users. Usually we have ${N > K}$, e.g., $N=128$ and
$K=16$ have been considered  \cite{1} \cite{2}. For signal
detection, the complex-valued system model can be directly converted
to a corresponding real-valued system model, then the estimate of
the ${2K \times 1}$ transmitted signal vector ${{\bf{\hat s}}}$
coming from $K$ difference users can be achieved by the classical
MMSE algorithm as \cite{2}
\begin{equation}\label{eq1}
{\bf{\hat s}} = \big ({{\bf{H}}^H}{\bf{H}} + {\sigma
^2}{{\bf{I}}_{2K}}\big)^{-1}{{\bf{H}}^H}{\bf{y}} = {{\bf{W}}^{ -
1}}\widehat {\bf{y}},
\end{equation}
where ${{\bf{H}}}$ is the ${2N \times 2K}$ MIMO channel matrix, which can be obtained through frequency-domain and/or time-domain
training pilots \cite{8}, ${{\sigma ^2}}$ is the additive white
Gaussian noise (AWGN) power, ${{{\bf{I}}_{2K}}}$  is an identity
matrix of size ${2K \times 2K}$, ${{\bf{y}}}$ is the ${2N \times 1}$
received signal vector at the BS, ${\widehat {\bf{y}} =
{{\bf{H}}^H}{\bf{y}}}$ can be interpreted as the matched-filter
output of ${{\bf{y}}}$, and finally ${{\bf{W}} ={{\bf{H}}^H}{\bf{H}}
+ {\sigma ^2}{{\bf{{\bf I}}}_{2K}}}$ denotes the MMSE filtering
matrix. Note that the direct computation of the matrix inversion
${{\bf{W}}^{ - 1}}$ requires relatively high complexity of ${{\cal
O}({K^3})}$.

\section{Proposed Signal Detection Based on Richardson Method}
Unlike the conventional (small-scale) MIMO systems with small number of
antennas, large-scale MIMO systems have a special property that the
MMSE filtering matrix ${{\bf{W}}}$ determined by the MIMO channel
matrix ${{\bf{H}}}$ is symmetric positive definite, which can be
proved as below.

\vspace*{+2mm} \noindent\textbf{Lemma 1}. {\it For signal detection
of large-scale MIMO systems, the MMSE filtering matrix ${{\bf{W}}}$
is symmetric positive definite}. \vspace*{+2mm}

\textit{Proof:} The column vectors of the channel matrix
${{\bf{H}}}$ in large-scale MIMO systems are asymptotically
orthogonal (i.e., ${{\rm{rank}}\left( {\bf{H}} \right) = 2K}$)
\cite{2}. Then we have the equation ${{\bf{H}\bf{r}} = 0}$ when and only when $\bf{r}$ is a ${2K \times 1}$ zero vector. Thus, for an
arbitrary non-zero ${2K \times 1}$ vector $\bf{r}$, we have
\begin{equation}\label{eq2}
{\left( {{\bf{H}\bf{r}}} \right)^H}{\bf{H r }} = {{\bf{r
}}^H}({{\bf{H}}^{H}}{\bf{H}}){\bf{r }} > 0,
\end{equation}
which indicates that the Gram matrix ${{\bf{G =
}}{{\bf{H}}^H}{\bf{H}}}$ is positive definite. In addition, as we
have ${{{\bf{G}}^H} = {({{\bf{H}}^H}{\bf{H}})^H} = {\bf{G}}}$, so
${{\bf{G}}}$ is also symmetric. Thus, the Gram matrix ${{\bf{G}}}$
is symmetric positive definite. Finally, as the noise power
${{\sigma ^2}}$ is positive, we can conclude that the MMSE filtering
matrix ${{\bf{W}} ={{\bf{H}}^H}{\bf{H}} + {\sigma ^2}{{\bf{{\bf
I}}}_{2K}}}$ is symmetric positive definite matrix, too.  \qed

The special property that the MMSE filtering matrix ${{\bf{W}}}$ in
large-scale MIMO systems is symmetric positive definite inspires us
to exploit the Richardson method \cite{7} to efficiently solve (1)
in an inversion-less way. The Richardson method is used to solve the
${N}$-dimension linear equation ${{\bf{Ax}} = {\bf{b}}}$, where
${{\bf{A}}}$ is the ${N \times N}$ symmetric positive definite
matrix, ${{\bf{x}}}$ is the ${N \times 1}$ solution vector, and
${{\bf{b}}}$ is the ${N \times 1}$  measurement vector. The
Richardson iteration can be described as
\begin{equation}\label{eq3}
{{\bf{x}}^{(i + 1)}} = {{\bf{x}}^{(i)}} + w \big({\bf{b}} -
{\bf{A}}{{\bf{x}}^{(i)}}\big),~i = 0, 1, 2, \cdot  \cdot  \cdot
\end{equation}
where the superscript ${i}$ denotes the number of iterations, and
${w}$ represents the relaxation parameter. Since ${{\bf{W}}}$ in (1)
is also symmetric positive definite as proved above, we can exploit
the Richardson method to estimate the transmitted signal vector
${{\bf{\hat s}}}$ without matrix inversion as below:
\begin{equation}\label{eq4}
{{\bf{s}}^{(i + 1)}} = {{\bf{s}}^{(i)}} + w\big(\widehat {\bf{y}} -
{\bf{W}}{{\bf{s}}^{(i)}}\big),~i = 0, 1, 2, \cdot  \cdot \cdot
\end{equation}
where the initial solution ${{{\bf{s}}^{(0)}}}$ can be usually set as a ${2K \times 1}$ zero vector without loss of generality due to no priori information of the final solution is available [7]. Such initial solution will not affect the convergence of the Richardson method, since the symmetric positive definite matrix ${{{\bf{W}}}}$ guarantees the convergence of the Richardson method for any initial solution as we will prove in the following Lemma 2. Consequently, the final accuracy will also not be affected by the initial solution if the number of iterations ${i}$ is large (e.g., ${i=5}$), as will be verified later in the simulation results. Since the relaxation
parameter  $w$ in (4) plays an important role in convergence, next
we will prove that the convergence of the Richardson method for any
initial solution when the relaxation parameter is appropriately
selected.

\vspace*{+2mm} \noindent\textbf{Lemma 2}. {\it For the
${N}$-dimension linear equation ${{\bf{Ax}} = {\bf{b}}}$, the
necessary and sufficient conditions for convergence of the
Richardson method is that the relaxation parameter ${w}$ satisfies
${0 < w < 2/{\lambda _1}}$, where ${{\lambda _1}}$ is the largest
eigenvalue of symmetric positive definite matrix ${{\bf{A}}}$}.
\vspace*{+2mm}

\textit{Proof:} We define ${{\bf{D}} = {{\bf{{\bf I}}}_N} -
w{\bf{A}}}$ and ${{\bf{c}} = w{\bf{b}}}$, where ${{\bf{D}}}$ is the
iteration matrix. Then the Richardson iteration (3) can be rewritten
as
\begin{equation}\label{eq5}
{{\bf{x}}^{(i + 1)}} = {\bf{D}}{{\bf{x}}^{(i)}} + {\bf{c}},~i = 0,
1, 2, \cdot  \cdot  \cdot
\end{equation}
We call the iteration procedure is convergent if ${\mathop {\lim }\limits_{i \to \infty } {{\bf{s}}^{(i)}} = {\bf{\hat s}}}$ and ${{\bf{\hat s}} = {\bf{B\hat s}} + {\bf{c}}}$ for any initial solution ${{{\bf{s}}^{(0)}}}$.

The spectral radius of iteration matrix ${{\bf{D}} \in
{\mathbb{R}^{N \times N}}}$ is the non-negative number ${\rho
({\bf{D}}) = \mathop {\max }\limits_{1 \le n \le N} \left| {{\mu
_n}({\bf{D}})} \right|}$, where ${{\mu _n}({\bf{D}})}$ denotes the
${n}$th eigenvalue of ${{\bf{D}}}$, and the necessary and sufficient
conditions for the convergence of (5) is that the spectral radius
should satisfy [7, Theorem 7.2.2]
\begin{equation}\label{eq6}
\rho ({\bf{D}}) = \mathop {\max }\limits_{1 \le n \le N} \left|
{{\mu _n}({\bf{D}})} \right| < 1.
\end{equation}
Without loss of generality, we use ${{\lambda _1} \ge {\lambda _2} \ge  \cdot  \cdot  \cdot  \ge {\lambda _N} > 0}$ to denote the ${N}$ eigenvalues of symmetric positive definite matrix ${{\bf{A}}}$, where ${{\lambda _1}}$ is the largest
one. Because of ${\bf{D}}= {\bf{I}}-w{\bf{A}}$, we have ${{\mu _n}({\bf{D}}) = 1 - w{\lambda _n}}$, where ${{\lambda _n}}$ is the ${n}$th eigenvalue of  ${{\bf{A}}}$, which can be substituted into
(6), and then we have ${0 < w < 2/{\lambda _1}}$.  \qed

\section{Computational Complexity}
The complexity in terms of required number of multiplications is
analyzed for comparison. It can be found from (4) that the ${i}$th
iteration of the proposed signal detection algorithm involves one
multiplication of a ${2K \times 2K}$ matrix ${{\bf{W}}}$  with a
${2K \times 1}$ vector ${{{\bf{s}}^{(i)}}}$, and one multiplication
of a constant relaxation parameter ${w}$ with a ${2K \times 1}$
vector ${\widehat {\bf{y}} - {\bf{W}}{{\bf{s}}^{(i)}}}$, thus the
required number of multiplications is ${4{K^{^2}} + 2K}$ for each
iteration.

Table~\ref{TAB1} compares the complexity of the conventional Neumann
series approximation algorithm \cite{6} and the proposed algorithm
based on Richardson method. It is well known that the complexity of the classical MMSE algorithm is ${{\cal O}({K^3})}$, and Table~\ref{TAB1} shows that the conventional Neumann
series approximation
algorithm can reduce the complexity from ${{\cal O}({K^3})}$ to ${{\cal O}({K^2})}$ when the
number of iterations is ${i = 2}$. However, the complexity is ${{\cal O}({K^3})}$ when ${i \geq 3}$. Since usually large value of $i$ is
required to ensure the final approximation performance (e.g., ${i =
5}$ as will be verified later by simulation results), the overall
complexity is still ${{\cal O}({K^3})}$, which indicates that only marginal
complexity reduction can be achieved. On the contrary, the complexity of the
proposed algorithm is reduced from  ${{\cal O}({K^3})}$
to ${{\cal O}({K^2})}$ for any arbitrary number of iterations.

\begin{table}[tp]
\caption{Computational Complexity} \label{TAB1}
\begin{center}
\begin{threeparttable}
\begin{tabular}{*{1}{L{0.7cm}}*{2}{L{3.3cm}}}
\toprule[1pt]
 & Conventional Neumann series approximation  \cite{6}& Proposed algorithm based on Richardson method \\
\hline \\ [-2 ex]
 ${i = 2}$ & ${12{K^2} - 4K}$ & ${8{K^2} + 4K}$\\
 ${i = 3}$ & ${8{K^3} + 4{K^2} - 2K}$  & ${12{K^2} + 6K}$\\
 ${i = 4}$ & ${16{K^3} - 4{K^2}}$  & ${16{K^2} + 8K}$\\
 ${i = 5}$ & ${24{K^3} - 12{K^2} + 2K}$  & ${20{K^2} + 10K}$\\
\toprule[1pt]
\end{tabular}
\end{threeparttable}
\end{center}
\end{table}

\section{Simulation Results}
The simulation results of the bit error rate (BER) performance against the signal-to-noise ratio (SNR) are provided
to compare the proposed signal detection algorithm with the recently proposed Neumann series approximation algorithm
\cite{6}, whereby the SNR is defined at the receiver. The BER performance of the classical MMSE algorithm with
complicated but exact matrix inversion is also included as the benchmark
for comparison. We consider an ${N \times K = 128 \times 16}$
large-scale MIMO system employing the modulation scheme of 64 QAM,
and the rate-1/2 convolutional code with ${[{133_o} \; {171_o}]}$
polynomial together with a random interleaver. We adopt flat Rayleigh fading channel. At the receiver, the
log-likelihood ratios (LLRs) are extracted from the detected signal
for soft-input Viterbi decoding \cite{9}. Through intensive simulations, we find out that when ${N}$ and ${K}$ are fixed, the largest eigenvalue of the MMSE filtering matrix ${{\bf{W}}}$ is around a certain value, and accordingly the relaxation parameter is set as $w=0.00645$ to guarantee the convergence.

\begin{figure}[h]
\setlength{\abovecaptionskip}{0pt}
\setlength{\belowcaptionskip}{3pt}
\begin{center}
\hspace*{-3mm}\includegraphics[width=0.9\linewidth]{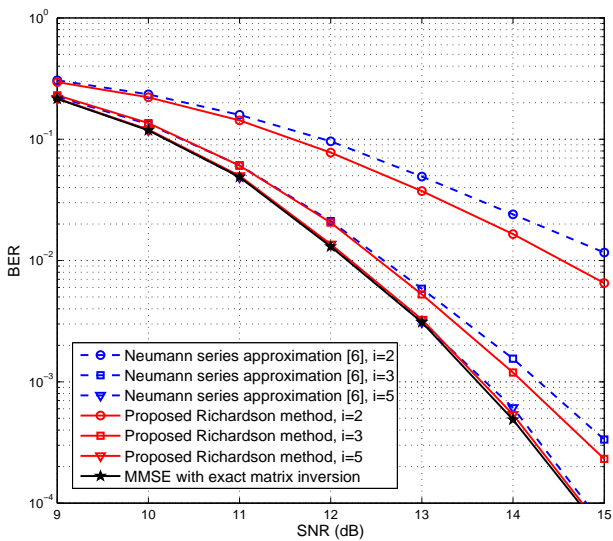}
\end{center}
\caption{BER performance comparison in an ${N \times K = 128 \times
16}$ large-scale MIMO system.} \label{FIG2}
\end{figure}

Fig. 1 shows the BER performance comparison results, where ${i}$
denotes the number of iterations. It is clear that the BER
performance of both algorithms improves with the number of
iterations, but the proposed algorithm outperforms the conventional
one when the same number of iterations is used, which indicates that
a faster convergence rate can be achieved by the proposed signal
detection algorithm. More importantly, when the number of iterations
is moderately large (e.g., ${i = 5}$ in Fig. 1), the proposed algorithm without
the complicated matrix inversion can achieve the near-optimal BER
performance of the MMSE algorithm with exact matrix inversion.


\section{Conclusions}
By fully exploiting the special property that the MMSE filtering
matrix in large-scale MIMO systems is symmetric positive definite,
we propose a low-complexity near-optimal signal detection algorithm based on the
Richardson method to avoid the complicated matrix inversion, which can
reduce the complexity from ${{\cal O}({K^3})}$ to ${{\cal
O}({K^2})}$. We also prove the convergence of the proposed algorithm
for any initial solution when the relaxation parameter is
appropriate. Simulation results verify that the proposed algorithm
outperforms the conventional method, and achieves the near-optimal
performance of the classical MMSE algorithm.

\vskip3pt \ack{This work was supported by National Key Basic
Research Program of China (Grant No. 2013CB329203), National Natural
Science Foundation of China (Grant No. 61201185), Science and Technology Foundation for Beijing Outstanding Doctoral Dissertation (Grant No. 2012T50093),
and the ZTE fund
project (Grant No. CON1307250001). }

\vskip5pt

\noindent Xinyu Gao, Linglong Dai, and Zhaocheng
Wang (\textit{Tsinghua National
Laboratory for Information Science and Technology, Department of
Electronic Engineering, Tsinghua University, Beijing 100084, China})

\noindent Yongkui Ma (\textit{Department of Electronic
Engineering, Harbin Institute of Technology, Harbin 150001, China})

\vskip3pt

\noindent E-mail: daill@tsinghua.edu.cn

\end{document}